\documentclass[journal=jacsat,manuscript=article]{achemso}

\usepackage{chemformula} % Formula subscripts using \ch{}
\usepackage[T1]{fontenc} % Use modern font encodings
\usepackage{hyperref}
\usepackage{url}
\usepackage{amsfonts}
\usepackage{xspace}

\newcommand\moldstruct{M\scalebox{0.8}{OL}DS\scalebox{0.8}{TRUCT}\xspace}

\author{Tomas André}
\affiliation[Uppsala University]
{Department of Physics and Astronomy, Uppsala University, Box 516, SE-751 20
Uppsala, Sweden}
\email{tomas.andre@physics.uu.se}

\author{Emiliano De Santis}
\affiliation{University of Rome Tor Vergata \& INFN, Rome, 00133, Italy}

\author{Nicu\c{s}or T\^\i mneanu}
\affiliation[Uppsala University]{Department of Physics and Astronomy, Uppsala University, Box 516, SE-751 20
Uppsala, Sweden}

\author{Carl Caleman}
\email{carl.caleman@physics.uu.se}
\affiliation[Uppsala University]{Department of Physics and Astronomy, Uppsala University, Box 516, SE-751 20
Uppsala, Sweden}
\alsoaffiliation[Second University]
{Center for Free-Electron Laser Science,
Deutsches Elektronen-Synchrotron, Notkestraße 85 DE-22607 Hamburg, Germany}

\title{Partial Orientation Retrieval of Proteins From Coulomb Explosions}

\abbreviations{}

\begin{document}

%%%%%%%%%%%%%%%%%%%%%%%%%%%%%%%%%%%%%%%%%%%%%%%%%%%%%%%%%%%%%%%%%%%%%
%% The "tocentry" environment can be used to create an entry for the
%% graphical table of contents. It is given here as some journals
%% require that it is printed as part of the abstract page. It will
%% be automatically moved as appropriate.
%%%%%%%%%%%%%%%%%%%%%%%%%%%%%%%%%%%%%%%%%%%%%%%%%%%%%%%%%%%%%%%%%%%%%
%\begin{tocentry}

%Some journals require a graphical entry for the Table of Contents.
%This should be laid out ``print ready'' so that the sizing of the
%text is correct.

%Inside the \texttt{tocentry} environment, the font used is Helvetica
%8\,pt, as required by \emph{Journal of the American Chemical
%Society}.

%he surrounding frame is 9\,cm by 3.5\,cm, which is the maximum
%permitted for  \emph{Journal of the American Chemical Society}
%graphical table of content entries. The box will not resize if the
%content is too big: instead it will overflow the edge of the box.

%This box and the associated title will always be printed on a
%separate page at the end of the document.

%\end{tocentry}

%%%%%%%%%%%%%%%%%%%%%%%%%%%%%%%%%%%%%%%%%%%%%%%%%%%%%%%%%%%%%%%%%%%%%
%% The abstract environment will automatically gobble the contents
%% if an abstract is not used by the target journal.
%%%%%%%%%%%%%%%%%%%%%%%%%%%%%%%%%%%%%%%%%%%%%%%%%%%%%%%%%%%%%%%%%%%%%
\begin{abstract}
Single Particle Imaging techniques at X-ray lasers have made significant strides, yet the challenge of determining the orientation of freely rotating molecules during delivery remains. In this study, we propose a novel method to partially retrieve the relative orientation of proteins exposed to ultrafast X-ray pulses by analyzing the fragmentation patterns resulting from Coulomb explosions. We simulate these explosions for 45 proteins in the size range 100 -- 4000 atoms using a hybrid Monte Carlo/Molecular Dynamics approach and capture the resulting ion ejection patterns with virtual detectors. Our goal is to exploit information from the explosion to infer orientations of proteins at the time of X-ray exposure. Our results demonstrate that partial orientation information can be extracted, particularly for larger proteins. Our findings can be integrated into existing reconstruction algorithms such as Expand-Maximize-Compress, to improve their efficiency and reduce the need for high-quality diffraction patterns. This method offers a promising avenue for enhancing Single Particle Imaging by leveraging measurable data from the Coulomb explosion to provide valuable insights about orientation.
\end{abstract}

%%%%%%%%%%%%%%%%%%%%%%%%%%%%%%%%%%%%%%%%%%%%%%%%%%%%%%%%%%%%%%%%%%%%%
%% Start the main part of the manuscript here.
%%%%%%%%%%%%%%%%%%%%%%%%%%%%%%%%%%%%%%%%%%%%%%%%%%%%%%%%%%%%%%%%%%%%%
\section{Introduction}

Proteins are often described as the building blocks of life due to their fundamental roles in virtually all biological processes across living organisms. Understanding these processes at the molecular level requires detailed knowledge of protein structures, as the structure of a protein is inherently linked to its function. Proteins perform a diverse array of functions, including enzymatic catalysis, signal transduction, and molecular transport, all of which are crucial for maintaining cellular functions. Currently, high-resolution protein structure is often determined through X-ray crystallography, which can resolve the structure to Ångström resolution~\cite{boutet2012high}. Crystallography has some drawbacks, particularly because many important proteins—especially membrane proteins, which constitute a large portion of drug targets~\cite{drug_target} and around 30\% of proteins in organisms~\cite{Bill2011} are very difficult or even impossible to crystallize due to their structure. This limits the types of sample that can be studied. Another issue is the crystal structure itself, as native protein structures in cells are not crystals, and the crystallization itself can alter the structure. A solution to this problem is Single Particle Imaging (SPI)~\cite{neutze2000potential} at X-ray Free Electron Lasers (XFEL), which can image non-crystalline samples as individual particles. SPI is now starting to mature and there have been significant advancements in the past decade both in sample delivery and analysis methods~\cite{ekeberg2024observation,yenupuri2024helium}, but improvement is needed to reach its full potential. Smaller proteins still pose a challenge and the smallest protein imaged with X-rays is 14~nm in diameter with around 60 000 atoms~\cite{ekeberg2024observation}. 

\begin{figure}
    \centering
    \includegraphics[width=0.9\linewidth]{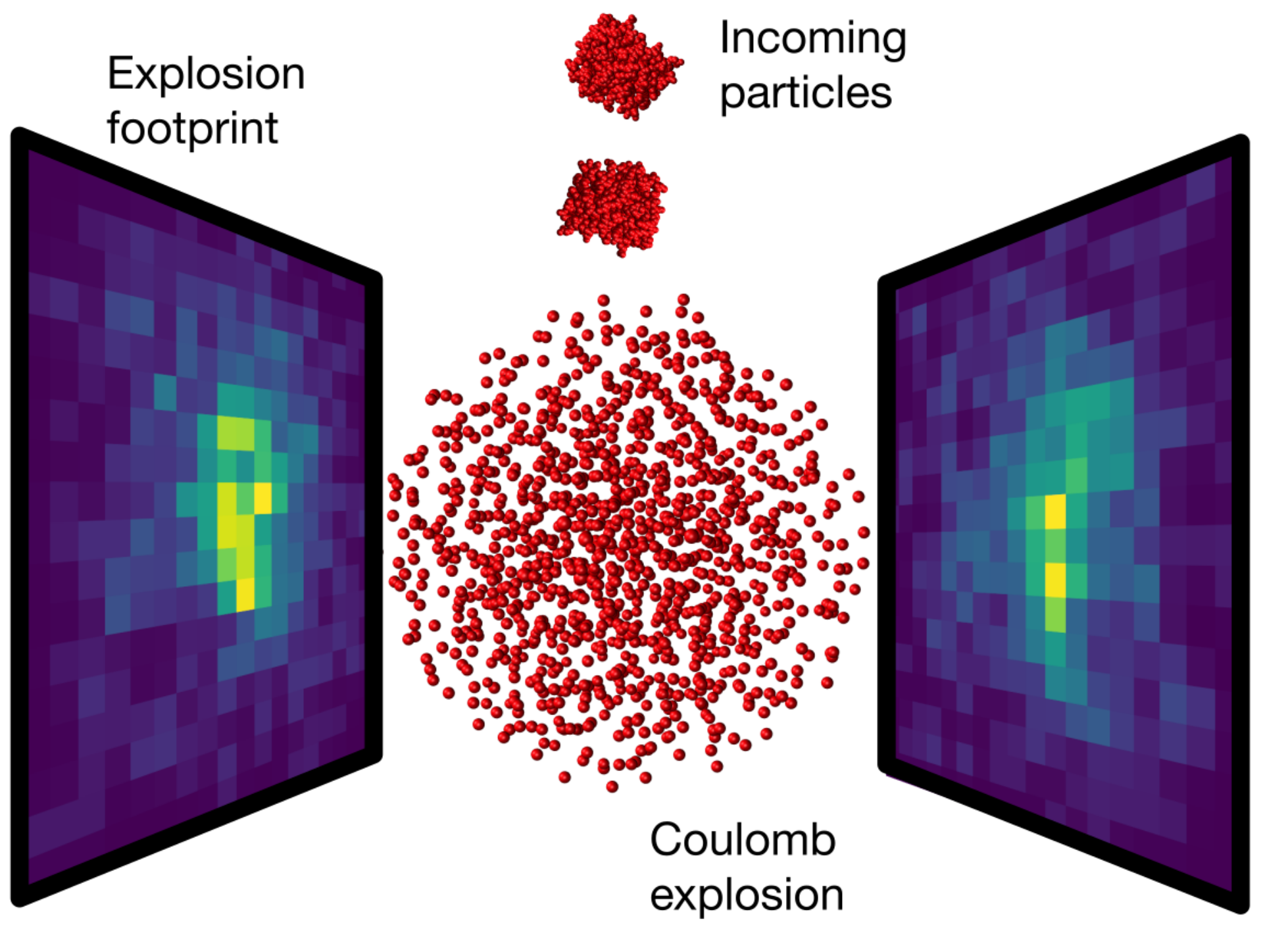}
    \caption{Geometry and setup of the simulated experiment. Detectors are placed on each side of the incoming particle at the XFEL interaction point. The detectors capture the ejected ions from the exploding sample. Perspective along X-ray beam.}
    \label{fig:detector}
\end{figure}

One of the challenges that remain to be overcome in SPI is the orientation of the sample~\cite{aquila2015roadmap}. During delivery each molecule can rotate freely in space, and there is currently no way of knowing this orientation. Using laser-induced alignment, there are ways to preorient small molecules~\cite{Alignment2014}, and there are studies investigating the possibility of preorienting larger molecules like proteins during delivery using electric fields~\cite{Marklund2017orientation,SINELNIKOVA2021}. T Algorithms such as Expand-Maximize-Compress (EMC)~\cite{loh_reconstruction_2009} can reconstruct a sample from diffraction data without any prior knowledge of its orientation; however, it requires many high-quality diffraction patterns.

Coulomb explosion imaging is a single-molecule structural determination technique for small molecules with around 10 atoms. In this technique, a sample is stripped of most of its electrons, and the fragments are tracked by measuring the momenta of the resulting ions in coincidence~\cite{vager_coulomb_1989}. XFELs provide a tool for carrying out Coulomb explosion imaging since X-rays can be tuned to target specific inner shells while reaching highly charged states via sequential single-photon absorption~\cite{kukk_molecular_2017,takanashi_ultrafast_2017,boll_x-ray_2022} and have the intensity to strip the electrons from the molecules. XFELs are powerful enough to cause a Coulomb explosion for protein macromolecules, although the number of ions makes measuring ions in coincidence unfeasible. However, \"Ostlin {\it et al.}~\cite{ostlin2018} simulated X-ray-induced Coulomb explosions on lysozyme to construct time-integrated \textit{explosion footprints} generated by projecting carbon and sulphur ions trajectories onto a virtual spherical detector and concluded these maps could be used to determine the protein's orientation during exposure if all ions can be detected. De Santis {\it et al.}~\cite{EDS2024} also studied how the spread of ions trajectories is influenced by at what depth within the protein they are located at during a Coulomb explosion.  Further, in a previous work we present the possibility of using explosion patterns to classify proteins solely based on the explosion footprint they create~\cite{andre2024}. Unlike conventional Coulomb explosion imaging used for very small molecules, we will consider a simplified approach, the explosion footprints used in this study are constructed uniquely from ion trajectories and carry no coincidence or momentum information.

In a recent study, Wollter \textit{et al.}~\cite{august2024} showed that partial knowledge about protein orientation can improve reconstruction capabilities of EMC as it would reduce the amount of  diffraction patterns required for successful reconstruction as well as allow the use of noisier patterns.

In the present theoretical work we ask the question \textit{Can we exploit the Coulomb explosion to get information about orientation?} Our goal is to describe a method for inferring partial orientation information that could aid the reconstruction process using EMC. This approach could increase the amount of usable data from measurements allowing for either shorter measurements or possibility to reconstruct samples which require more data.

\begin{figure*}
    \centering
    \includegraphics[width=\linewidth]{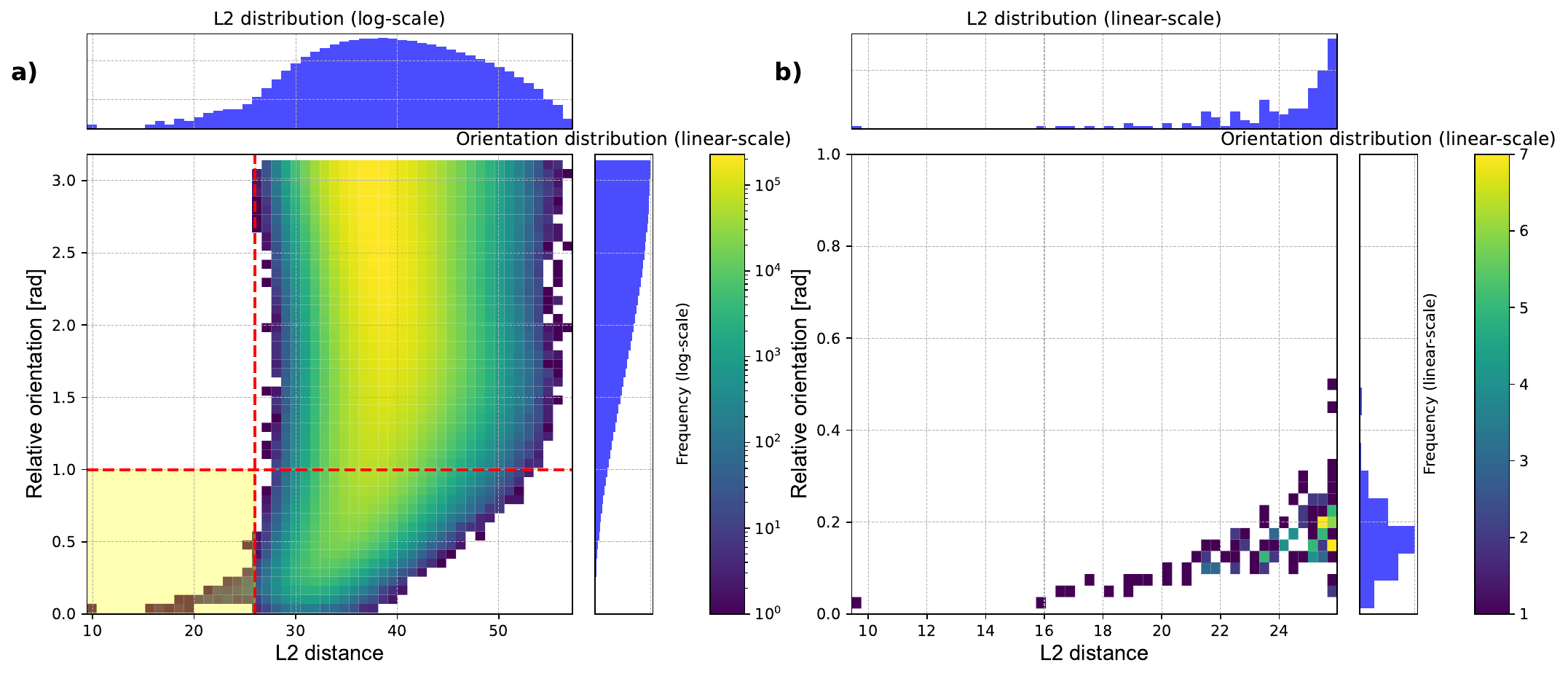}
    \caption{Panel a) Relative orientation between molecules against L2 distance (Eq.\ref{eq:L2}) between pairs of images from 10000 images from lysozyme. There is no global correlation but there is a trend for in the regime of lower L2 distance. We see from the distribution of points that the regime we are interested in contains a very small part of the data. Note that each point is a pair of images. Panel b) Zoomed in on the highlighted area of interest.}
    \label{fig:scatterplot}
\end{figure*}

\begin{figure*}
    \centering
    \includegraphics[width=\linewidth]{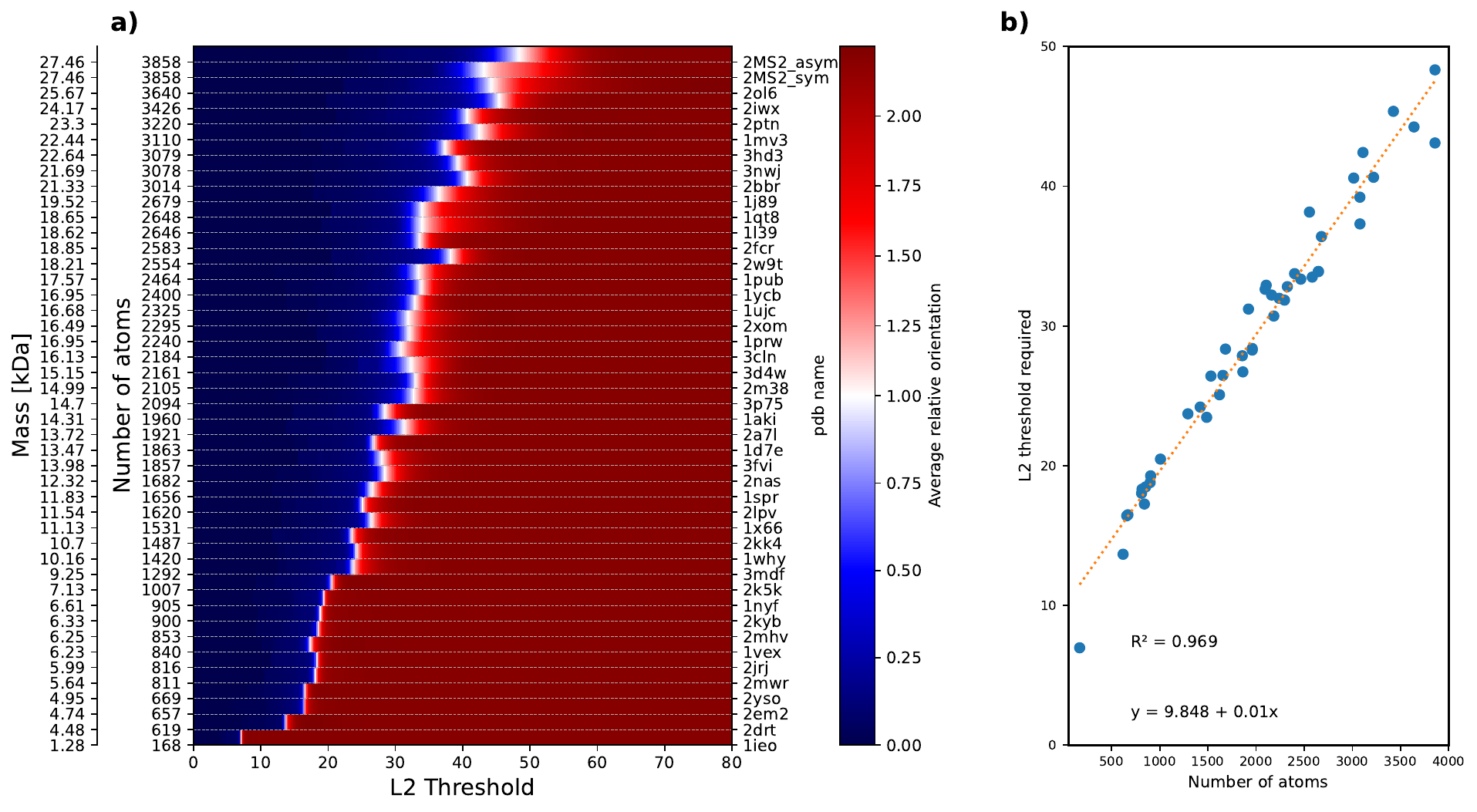}
    \caption{Panel a) The average relative orientation (colormap) of pairs under a given threshold (x-axis) for the set of 45 proteins simulated ordered by size from the bottom in ascending order. Note that the y-axis is not linear. Evaluated on 8000 images per protein. Where average relative orientation is 0 there is no data.
    Panel b) A linear regression on where the threshold is located plotted against the size of protein.}
    \label{fig:cutoff}
\end{figure*}

\section{Results}
With the image similarity metric (see the details of the method) and knowledge of the protein orientation from the simulations, the following section will show how we can infer the relative orientation of pairs of proteins from our virtual detector images. Using 10000 images for lysozyme (PdbID: \textit{1aki}) as an example, we can calculate the L2 distance (Eq. \ref{eq:L2}) between all pairs of images and then plot it against the ground truth pairwise relative orientation (Figure \ref{fig:scatterplot}). We see that for many patterns there are no correlations, which is expected since most orientations will not be close to each other when chosen at random. Pairs of images with small L2 difference correlates with small relative orientation between the proteins. So, by choosing a cutoff for the L2 distance, patterns with small relative orientation can be selected. We consider two orientations to be `close' if their relative orientation difference is less than 1 rad ($\approx 60^\circ$). Although this may seem like a large threshold, it is a meaningful choice. As demonstrated in Ref.~\cite{august2024}, any relative orientation information below 1 rad has been shown to improve reconstruction results when using EMC.
Smaller values can be chosen as well but in the following analysis we will use 1~rad. The location of the L2 cutoff is of course only known since we already know the orientations and can apply it by inspection. Following we will show how we can predict this cutoff based on the number of atoms in the sample and without any prior knowledge of the orientation. Noting the distributions of the relative orientations and the L2 values on the x- and y-axes in Figure \ref{fig:scatterplot}, we see that the region of interest contains a very small amount of all pairs of patterns, which statistically is not a surprise since by chance most proteins will not have very similar orientation. 

Our analysis of different proteins reveals that the required cutoff is not universal but is strongly influenced by the number of atoms in the protein. Figure \ref{fig:cutoff}a) illustrates the average relative orientation between all patterns for each protein under various cutoffs, with the colormap diverging at an average relative orientation of 1~rad. We note that, in general smaller proteins tend to require a smaller cutoff to achieve the required relative orientation. However, the number of atoms isn't the only factor influencing the cutoff values. Proteins with a similar number of atoms show a slight spread in their cutoff points, which likely is due to differences in their structural symmetries. These symmetries make the explosion footprint look very similar from different angles, affecting the cutoff values. For example, the patterns from a protein with 2-fold symmetry would be indistinguishable if we rotate it by \(\pi\) around the axis of symmetry.

\begin{figure}
    \centering
    \includegraphics[width=\linewidth]{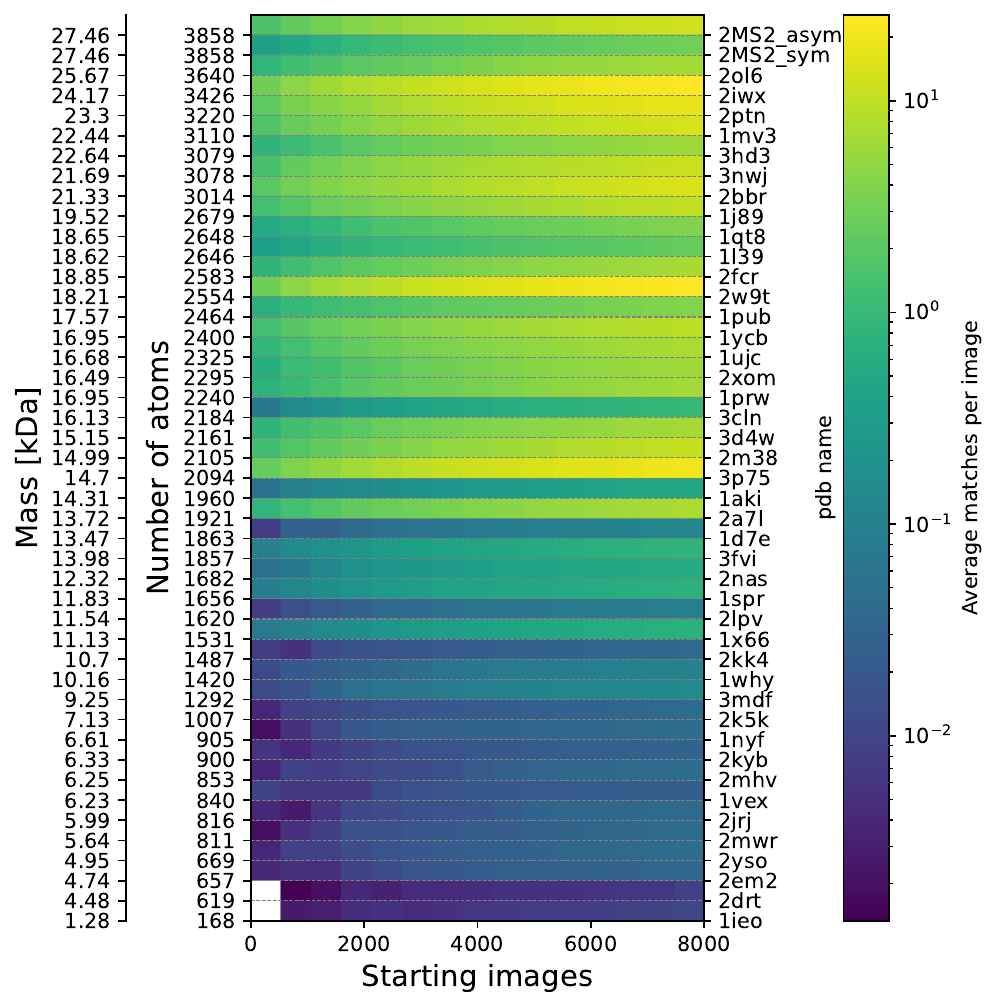}
    \caption{Average number matches (pairs under L2 threshold) per image as number of starting images increases. We expect more matches per image as the number of images is increased since we sample the space more finely and thus expect sampled orientations to be closer together on average.}
    \label{fig:matches}
\end{figure}

By finding the intercept along the L2 threshold where the average relative orientation is equal to 1 rad, we can find an estimate for what cutoff would be needed given a certain number of atoms. Figure \ref{fig:cutoff}b) shows this for the 45 proteins in this size range, the cutoff scales linearly with the number of atoms in the protein. There are, however some variations due to the different shapes and symmetries of the proteins.

The final step of our analysis is to determine how many pairs fall under the threshold. The number of matching pairs depends on the cutoff used; a smaller cutoff results in fewer matches but better orientation alignment, and the reverse is also true. To assess this, we count the number of matching pairs below the threshold for each protein, normalize it by the number of starting images, and then multiply by 2 (since each pair consists of two images). This allows us to determine the average number of neighbors each orientation has. The plot in Figure \ref{fig:matches} shows the number of matching neighbors for each protein across different numbers of starting images. We note that for larger proteins we find more matches, even with fewer starting images. As expected since we noted from Figure \ref{fig:cutoff} that we can choose a higher cutoff for the larger systems.

\section{Discussion}
\label{sec:discussion}
In this simulation study, we employed a Monte Carlo/Molecular Dynamics (MC-MD) model to investigate the novel use of Coulomb explosion patterns for retrieving partial orientation information of proteins during SPI experiments at XFELs. By simulating the Coulomb explosions and analyzing the resulting ion trajectories, our findings suggest that it is possible to infer relative orientations between protein pairs based on the explosion footprints. This approach offers a solution to the challenge of unknown protein orientations in SPI. We established a relationship between the number of atoms in a protein and the required L2 cutoff for determining orientation alignment. This provides a practical framework that could enhance reconstruction algorithms like EMC by complementing diffraction patterns with orientation information from ion patterns. Importantly, our simulations indicate that larger proteins, yield even better results with this method, likely due to the larger number of ions on the detector, reducing the apparent randomness for each ion path and ionization pathway. We hypothesize that this trend could extend also to proteins larger than those studied here. In 2011, the 400~nm Mimivirus became the first biological system ever imaged using SPI at an XFEL, marking a major milestone in the field~\cite{seibert2011}. Since then, the limits of SPI have been refined, allowing the imaging of smaller viruses and proteins, such as Photosystem II~\cite{suga2015native} and GroEL~\cite{ekeberg2024observation}, where the latter is only 14~nm, and is currently the smallest protein imaged with SPI. Although these proteins are larger than those considered in our study, we believe that incorporating ion data from Coulomb explosions could further lower this size limit. The partial orientation information could reduce the number of required diffraction patterns and allow for the inclusion of noisier data, making it a valuable tool for improving the efficiency and accuracy of SPI reconstructions.

In general, we observe that smaller proteins tend to require a lower cutoff, resulting in fewer matching images. The smallest protein in our study, a venom conopeptide from marine cone snails (PdbID: \textit{1ieo}), containing only 168 atoms, behaves as an outlier. The cutoff for this protein falls significantly lower than the predicted linear cutoff, indicating that the relationship between protein size and cutoff may not hold as well in the lower size range. Moreover, we obtained very few matches below the cutoff for smaller proteins, with noticeable improvement in the number of matches occurring around 2000 atoms, where each orientation get at least one neighbor on average, even for smaller amounts of starting images.

An interesting case involves the two largest proteins, a symmetric and an asymmetric dimer from the MS2 virus. Despite having the same number of atoms and identical amino acid sequences, the main differences between them are their distinct FG loops. For the symmetric structure, all FG loops are well defined $\beta$-hairpins, while for the asymmetric structure one of the FG loops is collapsed towards the main protein body~\cite{BRODMERKEL2022}. The symmetric structure requires a lower cutoff, demonstrating how protein symmetries can cause explosion footprints to overlap or appear indistinguishable across different orientations, thereby complicating the alignment process with the proposed method. In practice, symmetries should not pose any problem since the view along the symmetry orientations are equal.

The virtual detectors proposed could potentially be realized using a position-sensitive micro-channel plate detector. Such detectors already exist and have been used to detect similar patterns in smaller molecules~\cite{KORKULU2023232,boll_x-ray_2022}. A hypothetical experiment using such a detector appears plausible, as the parameters and geometry studied here are within experimental feasibility.
The detector presented in Ref.~\cite{KORKULU2023232} have the same dimensions as the virtual detector we use, but it has a resolution of around 500~\(\mu\)m. If we consider our pixels as channels in a similar detector, it would result in a resolution of about 6000~\(\mu\)m, around an order of magnitude smaller. A detector like this would most likely be easier to construct. Additionally, lower resolution improves the method we have presented here. This is for a similar reason as to why we apply the Gaussian kernel. With very small pixels, ions detected very close to each other but at different pixels other would not contribute to the similarity of the images, since we compute the distance between images pixel-wise. So in the case of using a detector with much higher resolution, it would be beneficial to bin the detector pixels into larger ones.

As mentioned in the introduction, partial orientation knowledge could aid in EMC reconstruction from diffraction patterns by allowing the use of noisier patterns. However, assessing how the number of matches improves reconstruction is beyond the scope of this paper, but it would be an interesting direction for future research.
Leveraging data from Coulomb explosions represents the next logical advancement for SPI, with the technology already available and the presence of ions in current experiments. Fundamentally SPI should work down to Ångström resolution, however the challenge still lies in refining methods like delivery systems, algorithms, XFEL brightness, and noise reduction to maximize the information extracted~\cite{aquila2015roadmap}. We believe that utilizing ion trajectories from Coulomb explosions will be key to achieving higher resolution for protein structures, even for smaller proteins than currently possible.

\section{Methods}
\subsection{Simulation setup}
All simulations are carried out using \moldstruct~\cite{dawod_moldstruct_2024,andre2024}\footnote{Available at \url{https://github.com/moldstruct/mc-md}}, a Monte~Carlo/Molecular~Dynamics code implemented in GROMACS~4.5.6~\cite{GROMACS}. The ionization model include photo-ionization, fluorescence, and Auger-Meitner decay. The model assumes all photons and electrons escape and therefore does not take electron collision effects into account. This is a reasonable assumption for samples in the size range we consider~\cite{dawod_moldstruct_2024}.
The simulations are initialized and equilibrated with the CHARMM36 forcefield~\cite{klauda2010update}. After energy minimization using steepest descent, we equilibrate in vacuum, at $300$~K, using a $1$~fs timestep, while keeping the protein's orientation fixed.  From this simulation, we sample structural variations in the protein that serve as initial starting points for the ionization simulations. In total we run $100$ simulations with different initial structures for each of the 45 different proteins. The proteins are sampled from the biological assemblies deposited in the Protein Data Bank in Europe (PDBe)\cite{velankar2016pdbe} within the range of $100$ -- $4000$ atoms, with the exception of two coat proteins of the MS2 virus (PdbId: \textit{2MS2}), which are dimers with the same amino acid sequence, however one is symmetric and one is asymmetric, we refer to them as \textit{MS2\_sym} and \textit{MS2\_asym} respectivly. These are hand-picked for their interesting symmetries. A full table of all proteins are given in supplemental information. We limit our study to proteins in this range as they are small enough to allow the free electrons to escape, which is a requirement for our simulation code to give accurate results. We assume particle injection speed is negligible compared to the speed of the ions during the Coulomb explosion, as typical injection speeds are on orders of $10^1$-$10^2$~m/s while the ions in our simulations are ejected at $10^{4}$-$10^{5}$~m/s. The Coulomb explosion simulations are carried out with a time step of $1$~as. The simulated XFEL pulse is Gaussian-shaped with a peak at $t=20$~fs and full width half maximum of $10$~fs. The pulse has a fluence of $5\cdot10^{6}$~photons/nm$^2$ and photon energy $2$~keV. These parameters are experimentally feasible at the SQS endstation at the European XFEL~\cite{EuXFEL_SQS_parameters}.The simulations run for 250~fs, after which the total energy in the systems consists of \(\approx 99\%\) kinetic energy, indicating that there is no more repulsion between the ions.

\subsection{Virtual detector}
By the end of the explosion simulations, the interaction dynamics between the ions have effectively ceased as they are now separated by large distances and continue to expand into space. From this point onward, their trajectories should follow relatively straight paths, allowing us to trace the direction of each ion to determine where it will end up. To capture the ions ejected from the Coulomb explosion, we envision placing two flat, square detectors, each with a side length of 120~mm, positioned 30~mm from the origin of the explosion on opposite sides, shown in Figure \ref{fig:detector}. We uniformly distribute $18\times18$ points on this surface, functioning as pixels for the detector. Motivation for this choice of detector is covered in the discussion.

By combining the images from both detectors, we create a new image represented as a \( (2\cdot18\times18)\) matrix. We extract the displacement vector of each ion and follow each ions path, each ion traced to the detector adds one count to the nearest bin, while ions falling outside the detector range are discarded. The resulting pattern, the explosion footprint, can be viewed as a 2D image, as seen on the face of the detectors in Figure \ref{fig:detector}.

\subsection{Orientation retrieval}
All simulations are carried out with the same orientation, in the analysis we will instead consider relative orientation between pairs of proteins. To do this we sample $N$ 3D-rotations uniformly as unit quaternions, which is common when numerically working with rotations, and apply them to our ion vectors. This has the effect of emulating simulations with different orientations. We augment our data by extracting multiple explosion patterns from each explosion simulation, at different orientations, As an example, if we extract $10000$ patterns in total from the \(100\) simulations, we take $10000/100 = 100$ patterns from each simulation, each pattern at a different orientation of the protein.  
We use the relative orientation between two proteins as the metric to monitor how close they are to each other. For a 3D object it is possible to decompose any set of rotations into as a single rotation by one angle around some axis,  thus for any pair of orientations, we can describe the difference in orientation with a single rotation instead of a composing rotation about the x-,y-, and z-axes.
For two proteins with orientation described by the quaternions \(R_i\) and \(R_j\), we can extract the relative orientation between the proteins from the unit quaternions as $\text{Re}[R_iR_j^{-1}] = a$, $\theta = 2\cos^{-1}(a)$. Repeating this for the $N$ patterns extracted, we can create $N(N-1)/2$ pairs of patterns, and calculate the relative orientation between the each pair of proteins.

To compare the patterns, we can measure the distance between pairs of images \(A,B\) in a \((2\cdot18\cdot18)\)-dimensional pixel space. Where the value of each pixel in an image correspond to its own dimension. For example a \((2\times2)\)-image \(I\), represented by the matrix \(I~=~\begin{bmatrix}
    0.0 & 2.0 \\
    1.0 & 3.0
\end{bmatrix}\), can be represented as the point \((0.0,2.0,1.0,3.0) \in \mathbb{R}^4\). The distance between those points can then be calculated using the Euclidean distance, also refereed to as L2 distance. The distance between \(A,B\) would then be given by 
\begin{equation}\label{eq:L2}
    \mathrm{L2}(A-B) = \sqrt{\sum_{i=1}^{n}(A_i-B_i)^2 },
\end{equation}
where \(n\) is the number of pixels in the pattern. Another technique we will use is to blur our patterns by applying a convolution with a 2D Gaussian kernel, \(G = \frac{1}{2\pi\sigma^2}\exp\left(-\frac{x^2+y^2}{2\sigma^2}\right)\) with $\sigma=0.5$. This will slightly smear out the patterns which is good to allow for some wiggle room when measuring similarity of images. In the case where the patterns look very similar, but the pixels which are hit are slightly shifted, our metric would rank that as low similarity. But by spreading out each hit over a few pixels we can overcome this problem.

%\newpage
%\subsection{List of simulated proteins}

%\\ *\tiny{Symetric and asymetric coat proteins from the MS2 virus with PdbID 2MS2.}
%\caption{PdbID, mass and number of atoms for each simulated systems. The list is sorted by mass in ascending order. All pdbs files were obtained from The RCSB Protein Data Bank \url{https://www.rcsb.org/}}
%\label{table:pdb_mass}
%\end{table}

%%%%%%%%%%%%%%%%%%%%%%%%%%%%%%%%%%%%%%%%%%%%%%%%%%%%%%%%%%%%%%%%%%%%%
%% The "Acknowledgement" section can be given in all manuscript
%% classes.  This should be given within the "acknowledgement"
%% environment, which will make the correct section or running title.
%%%%%%%%%%%%%%%%%%%%%%%%%%%%%%%%%%%%%%%%%%%%%%%%%%%%%%%%%%%%%%%%%%%%%
\begin{acknowledgement}
Project grants from the Swedish Research Council (2018-00740, 2019-03935, 2023-03900) are acknowledged, and the Helmholtz Association through the Center for Free-Electron Laser Science at DESY. 
EDS and CC acknowledge support from a Röntgen Ångström Cluster grant provided by the Swedish Research Council and the Bundesministerium für Bildung und Forschung (2021-05988). The computations were enabled by resources in projects NAISS 2023/22-1301, 2023/22-733 and 2024/5-140 provided by the National Academic Infrastructure for Supercomputing in Sweden (NAISS), funded by the Swedish Research Council through grant agreement no. 2022-06725.

\end{acknowledgement}

%%%%%%%%%%%%%%%%%%%%%%%%%%%%%%%%%%%%%%%%%%%%%%%%%%%%%%%%%%%%%%%%%%%%%
%% The same is true for Supporting Information, which should use the
%% suppinfo environment.
%%%%%%%%%%%%%%%%%%%%%%%%%%%%%%%%%%%%%%%%%%%%%%%%%%%%%%%%%%%%%%%%%%%%%
%\begin{suppinfo}
%Additional data available:
%\begin{itemize}
%  \item Simulation software \href{https://github.com/moldstruct/mc-md}{\moldstruct}.
%  \item Data available at \href{https://cxidb.org/id-231.html}{CXIDB}.
%\end{itemize}

%\end{suppinfo}

%%%%%%%%%%%%%%%%%%%%%%%%%%%%%%%%%%%%%%%%%%%%%%%%%%%%%%%%%%%%%%%%%%%%%
%% The appropriate \bibliography command should be placed here.
%% Notice that the class file automatically sets \bibliographystyle
%% and also names the section correctly.
%%%%%%%%%%%%%%%%%%%%%%%%%%%%%%%%%%%%%%%%%%%%%%%%%%%%%%%%%%%%%%%%%%%%%
\bibliography{references.bib}

\end{document}

%\newpage
\section{Additional data available}
\begin{itemize}
  \item Simulation software \href{https://github.com/moldstruct/mc-md}{\moldstruct}.
  \item Data available at \href{https://cxidb.org/id-231.html}{CXIDB}.
\end{itemize}

\section{List of simulated proteins}

\begin{table}[h!]
\centering
\scalebox{0.9}{
\begin{tabular}{|l|c|c|}
\hline
PDB & Mass (Da) & Number of Atoms \\
\hline
1ieo & 1281.57 & 168 \\
\hline
2drt & 4475.96 & 619 \\
\hline
2em2 & 4744.31 & 657 \\
\hline
2yso & 4954.43 & 669 \\
\hline
2mwr & 5639.56 & 811 \\
\hline
2jrj & 5992.95 & 816 \\
\hline
1vex & 6225.12 & 840 \\
\hline
2mhv & 6249.05 & 853 \\
\hline
2kyb & 6334.20 & 900 \\
\hline
1nyf & 6612.21 & 905 \\
\hline
2k5k & 7134.14 & 1007 \\
\hline
3mdf & 9251.43 & 1292 \\
\hline
1why & 10157.44 & 1420 \\
\hline
2kk4 & 10702.99 & 1487 \\
\hline
1x66 & 11126.42 & 1531 \\
\hline
2lpv & 11540.13 & 1620 \\
\hline
1spr & 11829.42 & 1656 \\
\hline
2nas & 12324.75 & 1682 \\
\hline
2a7l & 13722.83 & 1921 \\
\hline
3fvi & 13977.57 & 1857 \\
\hline
1d7e & 13471.14 & 1863\\
\hline
1aki & 14313.18 & 1960 \\
\hline
3p75 & 14704.02 & 2094 \\
\hline
2m38 & 14991.22 & 2105 \\
\hline
3d4w & 15146.57 & 2161 \\
\hline
3cln & 16126.61 & 2184 \\
\hline
2xom & 16493.41 & 2295 \\
\hline
1prw & 16553.04 & 2240 \\
\hline
1ujc & 16682.04 & 2325 \\
\hline
1ycb & 16954.43 & 2400 \\
\hline
1pub & 17574.27 & 2464 \\
\hline
2w9t & 18212.65 & 2554 \\
\hline
1l39 & 18623.48 & 2646 \\
\hline
1qt8 & 18645.51 & 2648 \\
\hline
2fcr & 18849.76 & 2583 \\
\hline
1j89 & 19524.71 & 2679 \\
\hline
2bbr & 21328.67 & 3014 \\
\hline
3nwj & 21692.14 & 3078 \\
\hline
1mv3 & 22435.82 & 3110 \\
\hline
3hd3 & 22638.11 & 3079 \\
\hline
2ptn & 23299.33 & 3220 \\
\hline
2iwx & 24173.62 & 3426 \\
\hline
2ol6 & 25674.60 & 3640 \\
\hline
2MS2$_{sym}$* & 27459.17 & 3858 \\
\hline
2MS2$_{asym}$* & 27459.17 & 3858 \\
\hline
\end{tabular}
}
\\ *\tiny{Symetric and asymetric coat proteins from the MS2 virus with PdbID 2MS2.}
\caption{PdbID, mass and number of atoms for each simulated systems. The list is sorted by mass in ascending order. All pdbs files were obtained from The RCSB Protein Data Bank \url{https://www.rcsb.org/}}
\label{table:pdb_mass}
\end{table}

%%%%%%%%%%%%%%%%%%%%%%%%%%%%%%%%%%%%%%%%%%%%%%%%%%%%%%%%%%%%%%%%%%%%%
%% The "Acknowledgement" section can be given in all manuscript
%% classes.  This should be given within the "acknowledgement"
%% environment, which will make the correct section or running title.
%%%%%%%%%%%%%%%%%%%%%%%%%%%%%%%%%%%%%%%%%%%%%%%%%%%%%%%%%%%%%%%%%%%%%

%%%%%%%%%%%%%%%%%%%%%%%%%%%%%%%%%%%%%%%%%%%%%%%%%%%%%%%%%%%%%%%%%%%%%
%% The same is true for Supporting Information, which should use the
%% suppinfo environment.
%%%%%%%%%%%%%%%%%%%%%%%%%%%%%%%%%%%%%%%%%%%%%%%%%%%%%%%%%%%%%%%%%%%%%
%\begin{suppinfo}
%Additional data available:
%\begin{itemize}
%  \item Simulation software \href{https://github.com/moldstruct/mc-md}{\moldstruct}.
%  \item Data available at \href{https://cxidb.org/id-231.html}{CXIDB}.
%\end{itemize}

%\end{suppinfo}

%%%%%%%%%%%%%%%%%%%%%%%%%%%%%%%%%%%%%%%%%%%%%%%%%%%%%%%%%%%%%%%%%%%%%
%% The appropriate \bibliography command should be placed here.
%% Notice that the class file automatically sets \bibliographystyle
%% and also names the section correctly.
%%%%%%%%%%%%%%%%%%%%%%%%%%%%%%%%%%%%%%%%%%%%%%%%%%%%%%%%%%%%%%%%%%%%%